\documentstyle[11pt,psfig,amstex]{article}
\addtocounter{secnumdepth}{1}
\newcommand{\x}{\text{\bf x}}

\newcommand{\q}{\text{\bf q}}

\newcommand{\r}{\text{\bf r}}
\newcommand{\eps}{\varepsilon}
\newcommand{\p}{\partial}
\newcommand{\dd}{\text{d}}
\newcommand{\bo}{\text{\bf 0}}
\newcommand{\k}{\text{\bf k}}
\numberwithin{equation}{section}
\begin{document}
\thispagestyle{empty}
\title{Global Persistence in Directed Percolation} 
\author{K. Oerding$^a$ and F. van Wijland$^b$\\\\
$^a$Institut f\"ur Theoretische Physik\\
Heinrich Heine Universit\"at\\
40225 D\"usseldorf, Germany\\\\
$^b$Laboratoire de Physique Th\'eorique et Hautes Energies$^1$\\
Universit\'e de Paris-Sud\\
91405 Orsay cedex, France}

\maketitle
\vspace{-1cm}
\begin{small}
\begin{abstract}
We consider a directed percolation process at its critical point. The 
probability that the deviation
of the global order parameter with
respect to its average has not changed its sign between 0 and $t$  decays with
$t$ as a power law. In space dimensions $d\geq 4$ the {\it  global persistence 
exponent} $\theta_p$ that
characterizes this decay is $\theta_p=2$ while for $d<4$ its
value is increased to first order in $\eps = 4-d$. Combining a method developed
by Majumdar and Sire~\cite{MajumdarSire} with renormalization group
techniques we compute the correction to
$\theta_p$ to first order in $\eps$. The global
persistence exponent is found to be a new and independent exponent. We finally 
compare our results with existing simulations. 

\end{abstract}
\end{small}
\vspace{1.5cm}
\noindent L.P.T.H.E. - ORSAY 98/26\\
\\\\\\\\\\{\small$^1$Laboratoire associ\'e au Centre National de la
Recherche Scientifique - URA D0063}
\newpage
\section{Motivations}
\subsection{Directed Percolation}
At the initial time $A$ particles are placed randomly with density 
$\rho_0$
on the sites of a $d$-dimensional hypercubic lattice. They perform independent simple random 
walks
with a diffusion contant $\lambda$. Multiple occupancy is allowed. The $A$ 
particles undergo three reaction processes: coagulation upon
encounter at a rate $k$, branching at a rate $k'$, spontaneous death at a rate 
$\gamma$, 
\begin{equation}\label{Schlogl}\begin{split}
A+A\stackrel{k}{\rightarrow}A\\
A\stackrel{k'}{\rightarrow}A+A\\
A\stackrel{\gamma}{\rightarrow}\emptyset
\end{split}\end{equation}
As the branching rate $k'$ is decreased below a threshold value $k_c^\prime$ (equal to
$\gamma$ in mean-field), the steady state of this system exhibits a
continuous transition from a state in which a finite  positive density of $A$'s
survive indefinitely to an absorbing $A$-free state. The order parameter of
the transition is $\rho_A$, the average of the local density of $A$'s.\\

We have used here the language of the Schl\"ogl reaction-diffusion process to
describe directed percolation as in \cite{Janssen}. Various alternative formulations exist
(\cite{CardySugar,GrassbergerSundermeyer}). Furthermore the scope of directed 
percolation reaches far beyond chemical kinetics,
as an overwhelmingly large class of nonequilibrium systems possessing a phase
transition in their steady state fall in the same universality class (cellular
automata, surface growth, reaction-diffusion processes). This makes of the 
process Eq.~(\ref{Schlogl}) a
paradigm for non-equilibrium systems with a transition in their steady
state. Our knowledge of the behavior of the system in the steady state and 
during
the relaxation stages rests on numerical simulations (in low space
dimensions, $d=1,2$) and on
analytical techniques (short time series expansions in $d=1,2$, renormalization 
group 
in $d=4-\eps$). The critical regime is characterized by a set
of three independent exponents: the dynamical exponent $z$, the anomalous
dimension of the order parameter $\eta$ and the correlation length exponent
$\nu$.
Scaling laws for $\rho_A$ can be extracted from special cases of 
\begin{equation}
\rho_A(t,k'-k_c^\prime,\rho_0)=b^{-\frac{d+\eta}{2}}{\cal
F}(b^{1/\nu}|k'-k_c^\prime|,b^{-z}t,b^{\frac{d-\eta}{2}}\rho_0)
\end{equation}
which holds in the limit $b\rightarrow\infty$ with  the arguments of $\cal
F$ fixed. Similar scaling relations exist for correlation functions.
\subsection{Global Persistence}
In this article we want to focus on a property that cannot be deduced from
the knowledge of the scaling properties of correlation functions alone. We first 
define the deviation
of the global time-dependent order parameter with respect to its average:
\begin{equation}\label{defPsi}
\Psi(t)\equiv\lim_{L\rightarrow\infty}L^{-d/2}\sum_{\x\in 
L^d}\Big[n_A(\x,t)-\langle n_A(\x,t)\rangle\Big]
\end{equation}
In Eq.~(\ref{defPsi}) we denote by $n_A(\x,t)$ the number of $A$
particles at site $\x$ at time $t$, in a particular realization of the
reaction-diffusion process. The brackets $\langle...\rangle$ denote an average with 
respect to
the set of microscopic realizations consistent with the initial conditions, the
rules Eq.~(\ref{Schlogl}) and the diffusion.\\

We define the
{\it global persistence} probability as the probability
that $\Psi$ remain of constant sign between $0$ and $t$. Similar quantities
have been considered (see \cite{MajumdarSire,Oerdingetal}) in critical
dynamics of magnetic systems, $\Psi$ simply
being the total magnetization. There it was shown that, following
a quench from a high temperature disordered state to the critical point, the 
global
persistence probability decays with time as a power law characterized  by a
universal exponent $\theta_p$. In critical dynamics the persistence probability 
is a quantity that appears naturally in the
description of the system while it relaxes to its 
equilibrium state. Our
motivation for the present work lies in the lack
of both qualitative and analytical picture of the onset of longe range
correlations in 
non-equilibrium systems relaxing to
their steady state. We believe that the knowledge of the global persistence
probability will shape our picture of the way the system organizes at
criticality.\\

The remainder of the article is divided as follows. We first recall in Sec.~2 the 
well known
correspondence between directed percolation and field theory. Following Majumdar
and Sire~\cite{MajumdarSire}, it is possible to obtain the global persistence probability from a 
careful analysis of the
autocorrelation of the global order parameter. This analysis is performed in
great detail in
Secs.~3 and 4. In Sec.~5 we turn to the
explicit calculation of the persistence exponent. In our
conclusion we compare our results with existing simulations.
\section{Field theoretic formulation}
There are several ways of mapping directed percolation onto a field
theory (\cite{Janssen,CardySugar}). The resulting field theory involves
a field $\psi$ whose average is the local density 
of $A$
individuals, and a conjugate field $\bar{\psi}$ ; dropping terms irrelevant in
the vicinity of the upper critical dimension $d_c=4$ the corresponding action reads
\begin{equation}\label{SDP}
S[\psi,\bar{\psi}]=\int
\dd^dx\;\dd t\;\Big[\bar{\psi}(\p_t+\lambda(\sigma-\Delta))\psi+\frac{\lambda
g}{2}\psi\bar{\psi}(\psi-\bar{\psi})-\rho_0\delta(t)\bar{\psi}\Big]
\end{equation}
The parameter $g$ can be expressed in terms of the original reaction rates
$k,k',\gamma$ and 
the mass in the propagator is
$\lambda \sigma=\gamma-k'$. The action Eq.~(\ref{SDP}) is
the starting point of the subsequent analysis. Renormalization group techniques
allow us to focus on scaling laws close to or at criticality, during the
relaxation process or in the steady state. From here on, as we shall eventually focus on
phenomena taking place at criticality, we set $\sigma=0$. We now summarize a few well-known
results on the renormalization of the action Eq.~(\ref{SDP}) that can be found 
{\it e.g.} in \cite{Bronzan}.\\
 
One first defines renormalized parameters and fields as follows
\begin{eqnarray}\label{defu}
\psi=\sqrt{Z}\psi_{\text{R}},\;\;
\bar{\psi}=\sqrt{Z}{\bar{\psi}}_{\text{R}},\;\;
\lambda=Z^{-1}Z_\lambda \lambda_{\text{R}},\;\;
\frac{g^2}{(8\pi)^{d/2}}=Z_\lambda^{-2}Z^{-1}Z_u u\mu^{\eps}
\end{eqnarray}
where $\mu$ is a momentum scale. From the one loop expression of the two and three-point vertex functions one 
deduces the values of the
$Z$-factors using dimensional regularization and the minimal subtraction scheme.
They read
\begin{eqnarray}
Z=1+\frac{u}{\eps}\label{Zpsi},\;\;\;
Z_\lambda=1+\frac{u}{2\eps},\;\;\;Z_u=1+\frac{8u}{\eps}
\end{eqnarray}
The $\beta$-function has the one-loop expression
\begin{equation}
\beta_u\equiv\mu\frac{\dd u}{\dd \mu}=
u(-\eps+2\gamma_\lambda+\gamma-\gamma_u)=u(-\eps+6u)
\end{equation}
where we have introduced the Wilson functions $\gamma_i\equiv\mu\frac{\dd\ln
Z_i}{\dd\mu}$, $i=\emptyset,\lambda,u$. The $\beta$-function has a stable nontrivial
fixed point $u^\star=\frac{\eps}{6}+{\cal O}(\eps^2)$. Critical exponents are
then obtained from linear combinations of the $\gamma_i(u^\star)$, {\it e.g.}
$z=2-\gamma_\lambda^\star+\gamma^\star$ and $\eta=\gamma^\star$.\\

We find it convenient to shift $\psi$ by 
its mean-field expression
\begin{equation}
\psi_{\text{mf}}(t)=\frac{\rho_0}{1+\frac{\lambda g}{2}\rho_0 t}
\end{equation}
Therefore the action expressed in terms of the fields
$\phi\equiv\psi-\psi_{\text{mf}}$ and $\bar{\phi}\equiv\bar{\psi}$ reads
\begin{equation}\begin{split}
S[\phi,\bar{\phi}]=&\int\Big[\bar{\phi}(\p_t+\lambda(\frac{g\rho_0}{1+
\frac{\lambda
g}{2}\rho_0t}-\Delta))\phi-\frac{\lambda g\rho_0}{2(1+\frac{\lambda
g}{2}\rho_0t)}{\bar{\phi}}^2\\&+\frac{\lambda 
g}{2}\phi\bar{\phi}(\phi-\bar{\phi})\Big]
\end{split}\end{equation}
We have thus gotten rid of the initial term localized at $t=0$. We will use the 
following notation: $G^{(n,m)}$ denotes the $(n+m)$-point correlation function
involving $n$ fields $\phi$'s and $m$ fields $\bar{\phi}$'s, as defined in
Eq.~(\ref{green}), and $W^{(n,m)}$ denotes its connected counterpart. The basic ingredients for a perturbative expansion are the free propagator $G$
and the free correlator $C$, defined by the zero-loop expression of $G^{(1,1)}$ 
and $G^{(2,0)}$, respectively. We shall need the large time behavior of $G$ and 
$C$:
\begin{equation}
G(\k;t',t)=\Theta(t'-t)\Big(\frac{t}{t'}\Big)^2\;\exp[-\lambda(\k^2+\sigma)(t'-t
)]
\end{equation}
\begin{equation}
\begin{split}
C(\k;,t,t')=\frac{2}{t_>^2t_<^2}\text{e}^{-\lambda\k^2(t_<+t_>)}
\frac{1}{(2\lambda\k^2)^4}\Big[&6(1-\text{e}^{2\lambda\k^2t_<})\\
&+6(2\lambda\k^2t_<)\text{e}^{2\lambda\k^2t_<}\\&-3(2\lambda\k^2t_<)^2\text{e}^{
2\lambda\k^2t_<}\\&+(2\lambda\k^2t_<)^3\text{e}^{2\lambda\k^2t_<}\Big]
\end{split}\end{equation}
in which we have set $t_<\equiv \text{min}\{t,t'\}$ and $t_>\equiv
\text{max}\{t,t'\}$. Note that the dependence on the initial density $\rho_0$
has disappeared, this is because we are focussing on times large with respect to
the time scale set by $\rho_0$. We have now in hands the building blooks for a
perturbation expansion of the expectation values of time-dependent observables.
\section{Autocorrelation function}
Our aim is to find the one-loop
correction to the function ${\cal C}(t,t')$ defined by
\begin{equation}
{\cal C}(t,t')\equiv \int \dd^dr\;\;
W^{(2,0)}(\r,t;\bo,t')=W^{(2,0)}(\k=\bo;t,t')
\end{equation}
which merely is
the autocorrelation function of the field $\psi$. In order to determine ${\cal C}(t,t')$ we carry out a perturbation expansion in powers of 
the coupling
constant $g$. The first term of this expansion is of course $C(\k=\bo;t,t')$. 
The first non
trivial corrections come in six pieces, each depicted by a one loop
connected Feynman diagram shown in Fig.\,1. The explicit calculation of these diagrams combined with
Eqs.\,(\ref{defu}) and (\ref{Zpsi}) allows one to determine the renormalized 
autocorrelation function
\begin{equation}
{\cal C}_{\text{R}}(t,t')=Z^{-1}{\cal C}(t,t')
\end{equation}
Using again the shorthand notations $t_<= \text{min}\{t,t'\}$,
$t_>= \text{max}\{t,t'\}$, we find
\begin{equation}\label{Cttfinal1}
{\cal C}_{\text{R}}(t,t')=\frac{1}{2}(\lambda t_<\mu^2)^{\frac{\eps}{12}}
\Big(\frac{t_<}{t_>}\Big)^{2-\frac{\eps}{4}}A\Big[1-\frac{\eps}{6}
F(t_>/t_<)+{\cal O}(\eps^2)\Big]
\end{equation}
which holds for $t_<,t_>$ large with $t_>/t_<$ finite. In
Eq.~(\ref{Cttfinal1}) the amplitude
$A$ reads
\begin{equation}
A=1+\frac{\eps}{6}\Big(\frac{1457}{600}+\frac{\pi^2}{20}-\frac{96}{25}\ln
2\Big)+{\cal O}(\eps^2)
\end{equation}
and the function $F$ has the expression
\begin{equation}
\begin{split}
F(x)=&-\frac{19}{30x}+\frac{11}{6}+\frac{\pi^2}{10}-\frac{96}{25}\ln 2
+\frac{12}{25}x-\frac{7}{50}x^2-\frac{77}{50}
x^3\\
&+\ln(1-x^{-1})\Big[\frac{101}{25}-5x+x^2+x^3-\frac{26}{25}x^4-\frac{1}{25}
x^5+\frac{1}{25x}\Big]\\
&+\ln(1+x^{-1})\Big[\frac{23}{50}+x+x^2+x^3+\frac{23}{50}x^4-\frac{1}{25}
x^5-\frac{1}{25x}\Big]\\
&+2\ln^2 x+3\ln x\ln(1-x^{-1})\\
&+4\text{Li}_2(1-x)+\text{Li}_2(1-x^{-1})-\frac{3}{20}(x^4-1)\text{Li}_2(x^{-2})
-\frac{3}{5}\text{Li}_2(x^{-1})
\end{split}
\end{equation}
where $\text{Li}_2(x)=-\int_0^x\dd t\ln(1-t)/t$ is the dilogarithm function. The limiting behavior of $F$ is found to be
\begin{equation}\begin{split}
F(\infty)=&\frac{8329}{1200}+\frac{2}{5}\pi^2+\frac{96}{25}\ln
2=0.3313...\\F(x)&\stackrel{x\rightarrow
1}{=}-2(x-1)\ln(x-1)+{\cal O}(x-1)
\end{split}\end{equation}
We have listed in the appendix the individual contributions to ${\cal
C}(t,t')$ arising from the corresponding Feynman diagrams.
\section{Short time expansion}
The result of the previous section Eq.~(\ref{Cttfinal1}) for ${\cal C}_{\text{R}}(t^{\prime}, t)$ holds for all times $t$
and $t'$, with $t/t'$ finite, but the limit $t\ll t'$ is singular. In this section we show that for $t\ll t^{\prime}$ the autocorrelation
function ${\cal C}_{\text{R}}(t^{\prime}, t)$ displays power law
behaviour with respect to
both time arguments and determine the corresponding exponents. In this
limit the random variable $\Psi(t)$ becomes a Markovian
process for which the persistence exponent may be expressed in terms of
well-known critical exponents. In the case of the Ising model the
Markovian approximation for $\theta_{p}$ is already close to the values
obtained by simulations~\cite{Majumdaretal,Oerdingetal}.

An appropriate method to study the correlation function for $t \ll
t^{\prime}$ is the short time expansion (STE) of the field $\psi({\bf
r}, t)$ in terms of operators located at the `time surface' $t=0$.
Since the Gaussian propagator and correlator are of the order
$t^{2}$ for $t \to 0$ we expect that the leading term in the STE is the
second time derivative $\ddot{\psi}$ of $\psi$, i.e.
\begin{equation}\label{STE}
\psi({\bf r}, t) - \langle \psi({\bf r}, t) \rangle = c(t) 
\ddot{\psi}({\bf r}, 0) + \ldots 
\end{equation}
(For $t=0$ the second time derivative of the response field $\bar{\psi}$
is equivalent to $\ddot{\psi}$.) The function $c(t)$ is a power of
$t$ which can be obtained from the difference of the scaling dimensions
of $\psi$ and $\ddot{\psi}$. Na\"{\i}vely, $c(t) \sim t^{2}$.

To compute the scaling dimension of $\ddot{\psi}$ in an
$\eps$-expansion one could determine the  additional renormalization
that is necessary to render correlation functions with $\ddot{\psi}$ insertions
finite. Fortunately, it is possible to express this
dimension to every order in $\eps$ in terms of other critical
exponents. For the initial density $\rho_{0} = \infty$ there is a
similarity between directed percolation and the semi-infinite Ising
model at the normal transition (i.e., for infinite surface
magnetization). In the latter case the short distance expansion of the
order parameter field near the surface is governed by the stress
tensor~\cite{Cardy,EKD}. Due to the translational invariance of the bulk
Hamiltonian the stress tensor requires no renormalization. Here we look
for an initial field which remains unrenormalized as a consequence of
the translational invariance (with respect to time) of the stationary
state.

Our argument applies to any dynamic field theory defined by a dynamic
functional of the form
\begin{equation}
S[\psi, \bar{\psi}] = \int_{0}^{\infty} \dd t \int \dd^{d}r \left(
\bar{\psi} \partial_{t} \psi - T[\psi, \bar{\psi}] \right)
\end{equation}
We assume that $\psi$ satisfies the sharp initial condition
$\psi({\bf r}, 0) = \rho_{0}$. For directed percolation we have
\begin{equation}\label{TDP}
T[\psi, \bar{\psi}] = -\lambda\bar{\psi}(\sigma-\Delta)\psi
-\frac{\lambda g}{2}\psi\bar{\psi}(\psi-\bar{\psi}) 
\end{equation}
Correlation functions may be written in the form
\begin{equation}
G^{(n,m)}(\{{\bf r}, t\}) = \int {\cal D}[\psi,\bar{\psi}]
\prod_{i=1}^{m} \bar{\psi}(\bar{{\bf r}}_{i}, \bar{t}_{i})
\prod_{j=1}^{n} \psi({\bf r}_{j}, t_{j}) \,\exp\left( -S[\psi,
\bar{\psi}] \right)
\label{green}
\end{equation}
where the functional integral runs over all histories $\{\psi,
\bar{\psi}\}$ which satisfy the initial condition.

We now introduce a new time variable $t \to t^{\prime} = t + a(t)$ (with
$\dot{a}(t) > -1$ to maintain the time order) and the transformed
fields $\bar{\psi}^{\prime}$ and $\psi^{\prime}$ with
$\bar{\psi}^{\prime}(t) = \bar{\psi}(t^{\prime})$ and $\psi^{\prime}(t)
= \psi(t^{\prime})$.
At lowest order in $a(t)$ the dynamic functional becomes
\begin{equation}
S[\psi, \bar{\psi}] = S[\psi^{\prime}, \bar{\psi}^{\prime}] -
\int \dd t \int \dd^{d}r \;\dot{a}(t) T[\psi^{\prime}, \bar{\psi}^{\prime}]
\end{equation}
where $a(0)=0$ has been assumed.

Performing the time shift in the correlation function $G^{(n,m)}$ and
comparing the terms of first order in $a(t)$ on both sides of
Eq.~(\ref{green}) one finds
\begin{equation}\begin{split}
\left(\sum_{i=1}^{m} a(\bar{t}_{i})
\frac{\partial}{\partial \bar{t}_{i}} + \sum_{j=1}^{n} a(t_{j})
\frac{\partial}{\partial t_{j}} \right) G^{(n,m)}(\{{\bf r}, t\})
 \\
= \langle \bar{\psi}(\bar{{\bf r}}_{1}, \bar{t}_{1}) \cdot
\ldots \psi({\bf r}_{n}, t_{n}) \int \dd t \int \dd^{d}r \dot{a}(t)
T[\psi, \bar{\psi}] \rangle 
\end{split}\end{equation}
Here the angular brackets 
indicate the average with respect to the weight
$\exp(-S[\psi, \bar{\psi}])$. We may choose
\begin{equation}
a(t) = a_{0} \left( 1 - {\rm e}^{-v t} \right)
\end{equation}
to obtain in the limit $v \to \infty$
\begin{equation}
\left(\sum_{i=1}^{m} \frac{\partial}{\partial \bar{t}_{i}} +
\sum_{j=1}^{n} \frac{\partial}{\partial t_{j}} \right)
G^{(n,m)}(\{{\bf r}, t\})
= \langle \bar{\psi}(\bar{{\bf r}}_{1}, \bar{t}_{1}) \cdot
\ldots \psi({\bf r}_{n}, t_{n}) \int \dd^{d}r T_{+} \rangle
\end{equation}
where $T_{+}$ denotes the operator $T[\psi, \bar{\psi}]$ in the limit
$t \to 0^+$. (We have assumed that all time arguments of the correlation
function are nonzero.)

This result shows that $T_{+}$ remains unrenormalized to every order of
the perturbation theory. Therefore its scaling dimension is given by
$d(T_{+}) = d + z$. At the upper critical dimension $d_{c}=4$ we
find $d(T_{+})=d(\ddot{\psi})=6$. In fact, one can show that
$T_{+}$ and $\ddot{\psi}(0)$ differ for $\rho_{0}^{-1}=0$ only by a
constant prefactor. To see this we express $T[\psi, \bar{\psi}]$ in terms of the
shifted field $\phi=\psi-\psi_{{\rm mf}}$. Since $\phi(t),\bar{\psi}(t)
\sim t^{2}$  for $t \to 0$ while $\psi_{{\rm mf}}(t) \sim
t^{-1}$ only the term $-(\lambda g/2)\psi_{{\rm mf}}^{2}\bar{\psi}$
contributes. Thus $T_{+} \sim \ddot{\bar{\psi}}(0) \sim \ddot{\psi}(0)$,
and the STE in Eq.~(\ref{STE}) becomes
\begin{equation}
\psi({\bf r}, t) - \langle \psi({\bf r}, t) \rangle = c(t) 
T_{+}({\bf r}) + \ldots 
\end{equation}
with
\begin{equation}
c(t) \sim t^{-(d(\psi)-d(T_{+}))/z} = t^{-((d+\eta)/2 - (d+z))/z} 
\end{equation}
Combining the STE with the general scaling form of the autocorrelation
function one obtains
\begin{equation} \label{STEcorr}
{\cal C}_{\text{R}}(t^{\prime}, t) \sim {t^{\prime}}^{-\eta/z} \left(
\frac{t}{t^{\prime}} \right)^{1 + (d-\eta)/(2z)} 
\end{equation}
which holds for $t/t'\rightarrow 0$.
\section{Global persistence}
\subsection{A detour via the Ornstein-Uhlenbeck process}
Let $X(\tau)$ be a Gaussian process with the following autocorrelation
function 
\begin{equation}\label{XXMarkov}
\langle X(\tau)X(\tau')\rangle=\text{e}^{-\theta_p^{(0)} |\tau'-\tau|}
\end{equation}
for $\tau,\tau'$ large (but arbitrary $\tau-\tau'$). The random variable $X$ is
thus a Gaussian stationary Markov process (of unit variance). It satisfies a Langevin equation
\begin{equation}
\frac{\dd X}{\dd\tau}=-\theta_p^{(0)} X(\tau)+\zeta(\tau),\;\;\;\theta_p^{(0)}>0
\end{equation}
where $\zeta$ is Gaussian white noise:
\begin{equation}
\langle\zeta(\tau)\zeta(\tau')\rangle=2\theta_p^{(0)}\delta(\tau-\tau')
\end{equation}
Therefore $X$ is an Ornstein-Uhlenbeck process. For such a process the probability that $X$ be positive between $0$ and $\tau$ 
decays exponentially as 
\begin{equation}
\text{Prob}\{\forall\tau'\in {[}0,\tau{]},
\;X(\tau')>0\}\propto \text{e}^{-\theta_p^{(0)}\tau}
\end{equation}
These are standard results.
\subsection{Expansion  around an Ornstein-Uhlenbeck process}
We now consider a Gaussian stationary process $X(\tau)$ which 
has the autocorrelation function
\begin{equation}\label{XXnonMarkov}
\langle X(\tau)X(\tau')\rangle=\text{e}^{-\theta_p^{(0)} |\tau'-\tau|}+\epsilon 
f(\tau'-\tau)
\end{equation}
with $f(0)=0$ and $\epsilon \ll 1$. Then $X$ is not a Markovian process. Majumdar and Sire~\cite{MajumdarSire} have shown how to evaluate the probability that $X$ be 
positive between $0$ and $\tau$ to
first order in $\epsilon$. They find
\begin{equation}
\text{Prob}\{\forall\tau'\in {[}0,\tau{]},
\;X(\tau')>0\}\propto \text{e}^{-\theta_p\tau}\end{equation}
where
\begin{equation}\label{thetapMS}
\theta_p=\theta_p^{(0)}\Big[1-\epsilon\frac{2\theta_p^{(0)}}{\pi}\int_0^\infty 
\dd \tau
\frac{f(\tau)}{(1-\exp(-2\theta_p^{(0)}\tau))^{3/2}}+{\cal O}(\epsilon^2)\Big]
\end{equation}
Hakim~\cite{Hakim} has extended this result to ${\cal O}(\epsilon^2)$.
\subsection{Application to the global order parameter}
At any fixed time $t$ there exists a dynamical correlation length $\xi\sim
t^{1/z}$ such that the system may be considered as a collection of effectively 
independent blocks
of linear size $\xi$. Hence $\Psi$ is the sum of $(L/\xi)^d$ independent degrees
of freedom, which is a Gaussian variable in the limit $L\rightarrow\infty$. We
are now in a position to apply the result Eq.~(\ref{thetapMS}) to the random 
variable
\begin{equation}
X(\tau)\equiv\Psi(\text{e}^{\tau})/\sqrt{\text{var}\Psi(\text{e}^\tau)}
\end{equation}
which has the autocorrelation function
\begin{equation}
\langle X(\tau)X(\tau')\rangle
=\text{e}^{-\theta_p^{(0)}|\tau'-\tau|}-\frac{\eps}{6}\;\text{e}^{-\theta_p^{(0)}|
\tau'-\tau|}F(\text{e}^{|\tau'-\tau|})
\end{equation}
with, after Eqs.~(\ref{STEcorr}) and (\ref{XXMarkov})
\begin{equation}
\theta_p^{(0)}=1+\frac{d}{2z}
\end{equation}
Substitution into Eq.~(\ref{thetapMS}) yields
\begin{equation}
\theta_p=\theta_p^{(0)}\Big[1+\frac{2\eps}{3\pi}{\cal I}+{\cal O}(\eps^2)\Big]
\end{equation}
where the integral
\begin{equation}
{\cal I}\equiv\int_1^\infty\dd x\frac{x^3
F(x)}{(x^4-1)^{3/2}}
\end{equation}
has the analytic expression
\begin{equation}
\begin{split}
{\cal I}=&\frac{13}{200}-\frac{9C}{20}-\frac{91\pi}{200}-
\frac{3\pi^2}{16}-\frac{9\pi}{80}\ln 2\\
&+\frac{\Gamma(\frac 14)^2}{\sqrt{2\pi}}\Big[\frac{\pi}{4}-\frac{1}{8}\ln
2+\frac{41}{80}\Big]
+\frac{\sqrt{2\pi^3}}{\Gamma(\frac
14)^2}\Big[-\frac{77\pi}{200}+\frac{23}{12}-\frac{3}{4}\ln 2\Big]\\
&+\frac{1}{50}
\;_3\!\text{F}_2(\frac 14,1,1;\frac 54,\frac 32;1)
-\frac{6}{5}\frac{\sqrt{2\pi^3}}{\Gamma(\frac 14)^2}
\;_3\!\text{F}_2(\frac 14,\frac 34,1;\frac 54,\frac 54;1)
\\
&-\frac{1}{30}\frac{\Gamma(\frac 14)^2}{\sqrt{2\pi}}
\;_3\!\text{F}_2(\frac 34,1,\frac 54;\frac 74,\frac 74;1)
+\frac{\sqrt{2\pi^3}}{\Gamma(\frac 14)^2}
\;_3\!\text{F}_2(\frac 12,\frac 34,1;\frac 54,\frac 32;1)
\\
&-\frac{1}{240}\frac{\Gamma(\frac 14)^2}{\sqrt{2\pi}}
\;_3\!\text{F}_2(\frac 14,1,1;\frac 74,2;1)
-\frac{1}{150}\frac{\sqrt{2\pi^3}}{\Gamma(\frac 14)^2}
\;_3\!\text{F}_2(\frac 34,1,\frac 32;\frac 94,\frac 52;1)\\
&+\frac{1}{2}\;_3\!\text{F}_2(\frac 34,1,1;\frac 32,\frac 74;1)
-\frac 38\frac{\Gamma(\frac 14)^2}{\sqrt{2\pi}}
\;_3\!\text{F}_2(\frac 14,\frac 12,1;\frac 34,\frac 32;1)\\
=&0.630237...
\end{split}
\end{equation}
where $C$ denotes Catalan's constant and $_3\text{F}_2$ the hypergeometric function of
order $(3,2)$. The final result reads
\begin{equation}
\theta_p=\theta_p^{(0)}(1+0.134\eps+{\cal O}(\eps^2))=2+0.059\eps+{\cal 
O}(\eps^2)
\end{equation}
\section{Discussion}
\subsection{Comparison to existing simulations}
Recently Hinrichsen and Koduvely~\cite{HinrichsenKoduvely} have performed a 
numerical study of one-dimensional
directed percolation in order to determine the asymptotic behavior of the global
persistence probability. In terms of the variable $\Psi$ defined in
Eq.~(\ref{defPsi}), they find the
following results. For the probability that $\Psi$ remain {\it negative} between
0 and $t$ they indeed find a power law decay characterized by a universal
exponent $\theta_p$ that has the numerical value $\theta_p=1.50(2)$. However
they find an exponential decay for the probability that $\Psi$ remain {\it
positive} between 0 and $t$. The latter assertion is in contradiction with our
finding that the global persistence probability decays algebraically {\it
irrespective} of the sign of $\Psi$. A plausible interpretation for such an
asymmetry could be the following. On the one hand the global persistence exponent is
well-defined in the regime in which the system has lost the memory of the
initial condition. This regime takes place for times $t$ such that
$t^{\frac{d-\eta}{2 z}}\rho_0\gg 1$. On the other hand, $\Psi$ is well defined
in the limit of infinitely large systems, and then it is the sum of a large number of effectively independent
contributions, which, on a lattice of size $L$, forces $L\gg \xi\sim t^{1/z}$.
Hence for numerical simulations to yield acceptable results, care must be taken
that the double limit $L^z\gg t\gg \rho_0^{-2z/(d-\eta)}$ is satisfied. Whether
the simulations in \cite{HinrichsenKoduvely} fulfill these bounds is 
questionable. Finally, in mapping the random  process $\Psi(t)$ to $X(\tau)$ we
have assumed that the time interval under consideration contains only times
large compared with $\rho_0^{-1}$ so that the regime in which $X$ is
stationary be reached. Strictly speaking, we should have defined the persistence probability 
over a time interval $[t_0,t]$, with $t\gg t_0\gg \rho_0^{-1}$. In a simulation
the choice $t_0=0$ leads to a persistence probability that enters the asymptotic regime
after times $t$ very large with respect to $\rho_0^{-1}$. This may be another problem with \cite{HinrichsenKoduvely}.

\subsection{Some speculations}
It is interesting to use the $\eps$-expansion to speculate on the numerical
value of $\theta_p$ in low space dimension. We define the {\it improved} value
of $\theta_p$, which we denote by $\theta_p^{\text{spec}}$, by the product of
the actual value of $\theta_p^{(0)}$ deduced from numerical
simulations, and the ${\cal O}(\eps)$ correction given by Eq.~(\ref{thetapMS}). 
We use recent simulations results of one and two-dimensional directed 
percolation carried out by Lauritsen {\it et al.}~\cite{Lauritsenetal}. 
\begin{equation}\label{recap}
\begin{array}{||c|c|c|c|c|c|c||}
\hline
d&z&\nu&\eta&\theta_p^{(0)}&\theta_p^{\text{spec}}&\theta_p\\
\hline
1&1.581&1.097&-0.496&1.316&1.8&1.50(2)\\
\hline
2&1.764&0.734&-0.409&1.567&2.0&-\\
\hline
4-\eps&2-\frac{\eps}{12}&\frac{1}{2}+\frac{\eps}{16}&-\frac{\eps}{6}&2-
\frac{5\eps}{24}&-&
2+0.059\eps\\
\hline
\geq 4&2&\frac{1}{2}&0&2&2&2\\
\hline
\end{array}
\end{equation}
In the first two lines the exponents $z,\nu$ and $\eta$ are taken from
\cite{Lauritsenetal} ; the value of $\theta_p^{(0)}$ was obtained 
using the hyperscaling relation $\theta_p^{(0)}=1+d/2z$, and that of $\theta_p$
in $d=1$ is taken from \cite{HinrichsenKoduvely}. Of course the column
$\theta_p^{\text{spec}}$ gives but a qualitative estimate of the true
$\theta_p$ that is supposedly closer to it than that obtained by the crude
$\eps$-expansion. These predictions certainly have to be tested against numerical
simulations.

\subsection{Final comments}
We would like to add some comments on table
(\ref{recap}). We find that $\theta_p>2$ in $d<4$ which says that both the
average and the variance of the time during which $\Psi$ is of constant sign are
finite. The average time depends on
a parameter which has the dimension of time, but there is no time
scale left in our treatment of the persistence probability. Therefore
the average time must depend on a microscopic scale or
on $\rho_{0}^{-1}$, which we have treated as a microscopic scale.
This may also explain why the average time is infinite for critical
dynamics: For zero initial magnetization a time scale such as $\rho_{0}^{-1}$
does not exist. In critical dynamics one could also define a persistence exponent for the critical relaxation
from an initial state with nonzero local magnetization. In that
case the persistence exponent would read in the Markovian approximation
$\theta_p^{(0)} = 1+d/(2z) > 1$ (as for the problem we have treated in this
article). \\\\

\noindent {\bf Acknowledgments}\\

F.v.W. would like to thank C. Sire for a discussion on
\cite{MajumdarSire}, V. Hakim for communicating \cite{Hakim}, and H.J.
Hilhorst for his interest and critical
comments. The work of K.~O. has been supported in part by the
Sonderforschungsbereich 237 [Unordnung und Gro{\ss}e Fluktuationen
(Disorder and Large Fluctuations)] of the Deutsche
Forschungsgemeinschaft.

\section{Appendix}
We introduce the notations $t_<\equiv \text{min}\{t,t'\}$ and $t_>\equiv
\text{max}\{t,t'\}$ and the ratio $r\equiv t_>/t_<$. The one loop contributions
to the function ${\cal C}(t,t')$ are the following.
\begin{equation}\begin{split}
\text{Fig.~1(a)}=\int \dd \tau\dd\tau' \Big[C(\bo;\tau,t)\int\frac{\dd^d q}{(2\pi)^d}
G(\q;\tau';\tau)^2 G(\bo;t',\tau')\\+C(\bo;\tau,t')\int\frac{\dd^d q}{(2\pi)^d}
G(\q;\tau';\tau)^2 G(\bo;t,\tau')\Big]
\end{split}\end{equation}
\begin{equation}
\begin{split}
\text{Fig.
1(a)}=\frac{g^2\mu^{-\eps}}{(8\pi)^{d/2}}r^{-2}(\mu^2 \lambda 
t_<)^{\eps/2}&
\Big[\frac 32 \frac{1}{\eps}+\frac{1}{\eps}\ln r-\frac
15\frac{1}{r}-\frac{21}{80}-\frac 18 r-\frac{3}{16}r^2-\frac{3}{8}r^3\\
&-\frac{3}{8}(r^4-1)\ln(1-r^{-1})\\&+\frac 12\ln
r\ln(1-r^{-1})+\frac 14\ln^2 r\\&-\frac 
12\text{Li}_{2}(1-r^{-1})\Big]\end{split}
\end{equation}
\begin{equation}
\begin{split}
\text{Fig.
1(b)}=\frac{g^2\mu^{-\eps}}{4(8\pi)^{d/2}}r^{-2}(\mu^2 \lambda 
t_<)^{\eps/2}&
\Big[\frac{551}{150}-\frac{\pi^2}{10}-\frac 13
\frac{1}{r}-\frac{169}{150}r+\frac{3}{100} r^2+\frac{129}{50}
r^3\\&+\ln(1-r^{-1})\Big(\frac{21}{50}+2r-2r^2-2r^3\\&+\frac{79}{50}
r^4+\frac{2}{25}
r^5-\frac{2}{25}\frac{1}{r}\Big)\\&+\ln(1+r^{-1})\Big(-\frac{23}{25}-2r-2
r^2-2r^3\\&-\frac{23}{25} r^4+\frac{2}{25}
r^5+\frac{2}{25}\frac{1}{r}\Big)\\&+\frac{3}{10} (r^4-1)\text{Li}_2(r^{-2})
+\frac{6}{5}\text{Li}_2(r^{-1})\Big]\end{split}
\end{equation}
\begin{equation}\begin{split}
\text{Fig.
1(c)+(d)}=\frac{g^2\mu^{-\eps}}{(8\pi)^{d/2}}r^{-2}(\mu^2 \lambda 
t_<)^{\eps/2}&
\Big[-\frac{1}{\eps}+\frac{2}{\eps}\ln
r\\&-\frac{43}{120}+\frac{3}{5}\frac{1}{r}+\frac{1}{6}r+\frac 14 r^2+\frac{1}{2}
r^3\\&-\ln(1-r^{-1})(\frac 52-2r-\frac 12 r^4)\\&-\frac 12 \ln^2 r-\frac 32 \ln
r-2\ln r\ln(1-r^{-1})\\&-2 \text{Li}_2(1-r)\Big]
\end{split}
\end{equation}
\begin{equation}
\text{Fig.
1(e)+(f)}=\frac{g^2\mu^{-\eps}}{(8\pi)^{d/2}}r^{-2}(\mu^2 \lambda 
t_<)^{\eps/2}
\Big[-\frac{3}{\eps}\ln r+\frac{9}{4}\ln r-\frac{3}{4}\ln^2 r\Big]
\end{equation}

\newpage
$$\text{\bf FIGURES}$$
\begin{figure}[hp]
\centerline{\psfig{figure=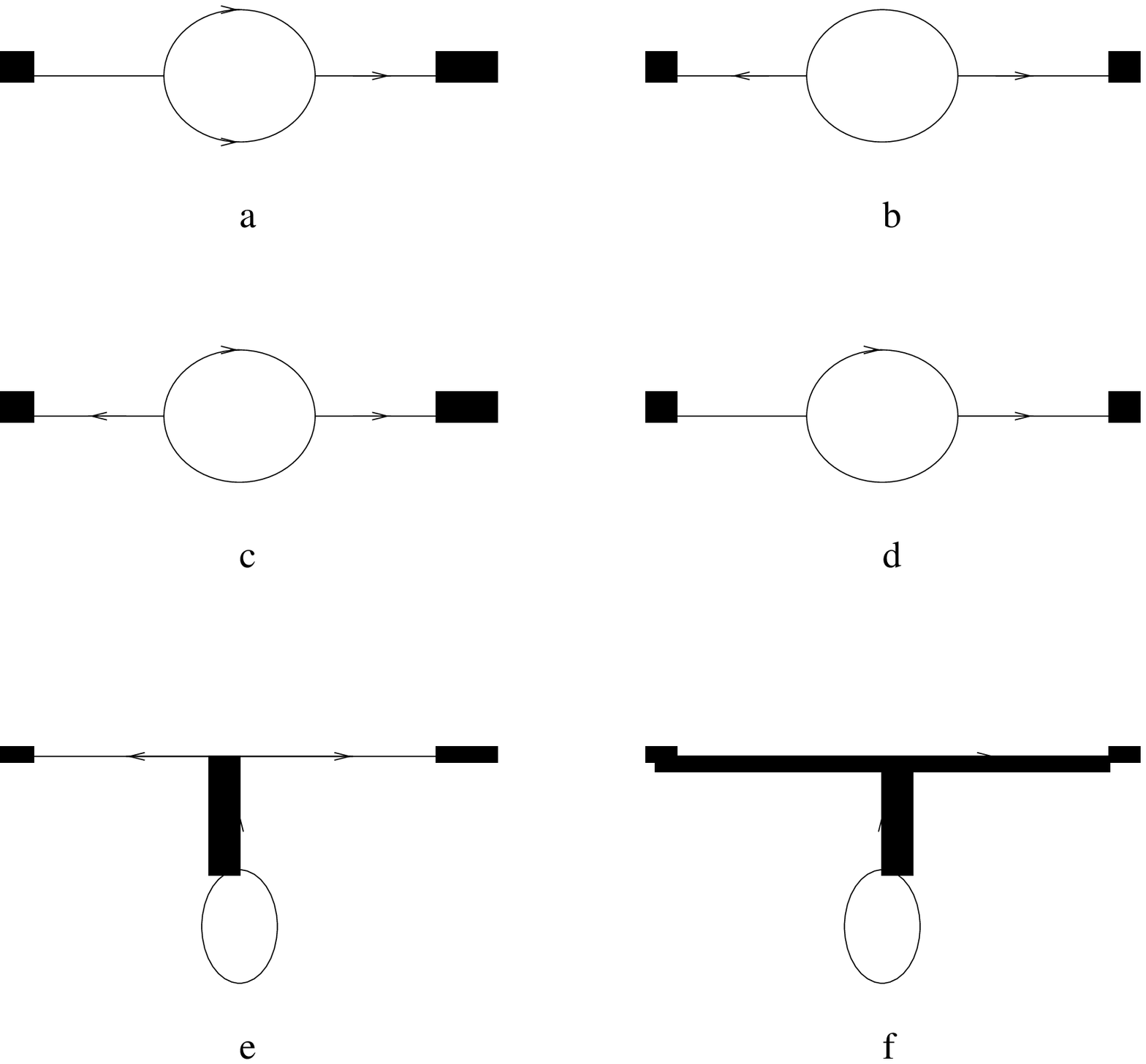,height=9cm,angle=0}}
\medskip
\caption{\small One-loop diagrams involved in the expression of ${
\cal C}(t,t')$. Note that diagrams (a,d,f) are the only one loop contributions
to $\langle\ddot{\psi}(\q=\bo,0)\psi(\q=\bo,t)\rangle$.}
\end{figure}

\end{document}